\documentclass[a4paper,article,12pt,showkeys]{revtex4-1}
\usepackage{amsmath,amssymb,times}
\usepackage{amsthm,amscd,ulem,bm}

\newtheorem{lemma}{Lemma}
\newtheorem{theorem}[lemma]{Theorem}

\begin{document}

\title{On Metrizability of Invariant Affine Connections}

\author{Erico Tanaka}

\affiliation{Faculty of Science, Palacky University, 17. listopadu 12\\
Olomouc, 771 46, Czech Republic} 
\email{erico.tanaka01@upol.cz} 

\affiliation{Advanced Research Institute for Science and Engineering,
Waseda University, 3-4-1 Ohkubo, Shinjuku, Tokyo, Japan }

\author{Demeter Krupka}

\affiliation{School of Mathematics, Beijing Institute of Technology,\\
No.5 South Zhongguancun Street, Beijing, 100081, China}
\email{demeter.krupka@lepageri.eu} 

\affiliation{Faculty of Science, The University of Ostrava, 30. dubna 22\\
Ostrava, 70103, Czech Republic}

\affiliation{Department of Mathematics, La Trobe University, Melbourne, Bundoora
Victoria, 3086, Australia }

\begin{abstract}
The metrizability problem for a symmetric affine connection on a manifold, invariant with respect to a group of diffeomorphisms $G$, is considered. We say that the connection is $G$-metrizable, if it is expressible as the Levi-Civita connection of a $G$-invariant metric field. In this paper we analyze the $G$-metrizability equations for the rotation group $G = \operatorname{SO} (3)$, acting canonically on three- and four-dimensional Euclidean spaces. We show that the property of the connection to be $\operatorname{SO} (3)$-invariant allows us to find complete explicit description of all solutions of the $\operatorname{SO} (3)$-metrizability equations. 
\end{abstract}

\keywords{affine connection; metrizability; $G$-invariant.}
\maketitle
\section{Introduction}

Let $g$ be a metric field on an $n$-dimensional manifold $X$, i.e., a non-singular symmetric tensor field of type $(0,2)$, and let ${}^g\Gamma $ be the {\it Levi-Civita connection} of $g$. 
We know that if $g$ has in a chart $(U,\varphi )$, $\varphi  = ({x^i})$, an expression $g = {g_{ij}}d{x^i} \otimes d{x^j}$, then the components of ${}^g\Gamma $ are the {\it Christoffel symbols} ${}^g\Gamma _{jk}^i$, defined by the decomposition 
\begin{eqnarray}
\frac{{\partial {g_{ij}}}}{{\partial {x^k}}} = {g_{is}}{}^g\Gamma _{jk}^s + {g_{js}}{}^g\Gamma _{ik}^s. \label{eq1}
\end{eqnarray}
Now let $\nabla $ be a symmetric affine connection on $X$. Recall that $\nabla $ is said to be {\it metrizable}, if 
\begin{eqnarray}
\nabla  = {}^g\Gamma \label{eq2}
\end{eqnarray}
for some metric field $g$. The {\it metrizability problem} for $\nabla $ consists in finding integrability conditions and solutions $g$ of this equation. If $\nabla _{jk}^i$ are the components, then the {\it local metrizability problem} consists of solving the system of partial differential equations 
\begin{eqnarray}
\frac{{\partial {g_{ij}}}}{{\partial {x^k}}} = {g_{is}}\nabla _{jk}^s + {g_{js}}\nabla _{ik}^s \label{eq3}
\end{eqnarray}
for unknown functions ${g_{ij}}$ such that ${g_{ij}} = {g_{ji}}$ and $\det {g_{ij}} \ne 0$. Equations (\ref{eq3}) are the {\it local metrizability conditions}. 

The metrizability problem has been studied by many authors, with different results and ideas (Anastasiei~\cite{ref1}, Crampin~\cite{ref2}, Crampin, Prince and Thompson~\cite{ref3}, Kowalski~\cite{ref7,ref8}, Krupka and Sattaro~\cite{ref9}, Sarlet~\cite{ref12}, Schmidt~\cite{ref13}, Tamassy~\cite{ref14}, and others). Our aim in this paper will be to follow straightforward approach to the problem, initiated by Eisenhart and Veblen~\cite{ref5}, who analyzed the system (\ref{eq2}) directly and derived some necessary and sufficient conditions for existence of its solutions. A general feature of all these results is the absence of explicit formulas for metrizable connections or the corresponding metric fields. On the other hand, many examples of metrizable and non-metrizable affine connections are known and can be found in Vilimova~\cite{ref16}. 

In application to physics, for example, the metrizability problem arises when considering Hamilton equations of general relativity, where independent field variables are a metric and a connection (see Krupka and Stepankova~\cite{ref10}). Considerations of such problems may in principle be utilized in finding solutions to the Einstein equations. 

Our new idea in this paper is to analyze the metrizability problem for affine connections, obeying certain invariance properties. We wish to clarify whether these additional properties could simplify the metrizability problem and lead to explicit solutions of metrizability equations.

Thus, suppose that $X$ is endowed with a left action of a Lie group $G$, and we have a {\it G-invariant} affine connection $\nabla $ on X. Then the {\it G-metrizability problem} consists in finding {\it G-metrizability conditions} for $\nabla $ that assure existence of a $G$-invariant metric field $g$ whose Levi-Civita connection is $\nabla $. 

Sometimes it is useful to formulate the g-metrizability in terms of vector fields; the corresponding definitions apply to one-parameter subgroups of $G$. If $\xi $ is a vector field on $X$, then we say that the metric field $g$ (resp. the affine connection $\nabla $) on $X$ is $\xi $-invariant, if the Lie derivative ${\partial _\xi }g$ (resp. ${\partial _\xi }\nabla $) vanishes, i.e., ${\partial _\xi }g = 0$ (resp. ${\partial _\xi }\nabla  = 0$). 
We say that a $\xi $-invariant affine connection $\nabla $ is $\xi ${\it -metrizable}, if there exists a $\xi ${\it -invariant} metric field $g$ such that condition (\ref{eq2}) is satisfied. Thus, we have the system of $\xi ${\it -metrizability equations} 
\begin{eqnarray}
{\partial _\xi }g = 0,\quad {}^g\Gamma  = \nabla . \label{eq4}
\end{eqnarray}

One can also say in terminology used by Prince and Crampin~\cite{ref11} that the $\xi $-metrizability problem is the metrizability problem in which the transformations, belonging to the one-parameter group of $\xi $ are {\it affine collineations} of the affine connection $\nabla $ and {\it isometries} of the metric field $g$. 

The notational conventions and a summary of basic concepts are set out in Section 2. Section 3 is devoted to $\xi $-metrizability and the structure of the $\xi $-metrizability equations. In Sections 4 and 5 we analyze two examples. We consider the canonical left actions of the rotation group $G = \operatorname{SO} (3)$ on the manifold $X = {R^3}\backslash \left\{ {(0,0,0)} \right\}$ and on $X = R \times ({R^3}\backslash \left\{ {(0,0,0)} \right\})$, diffeomorphic with ${R^2} \times {S^2}$. We find in these cases explicit solutions of the system (\ref{eq4}). 

The proofs of our theorems are omitted since they can be reconstructed from the formulas given in the text and from formulation of the theorems. 

Possible applications of our results and the method we present could be used in the geometry of affine connections and in physical field theories that include an affine connection as an independent field variable.

\section{Invariant Metrizability Problem}

The purpose of this section is to formulate the invariant metrizability problem for affine connections on a given smooth $n$-dimensional manifold $X$. We begin with basic criteria and infinitesimal invariance criteria for $(0,2)$-tensor fields and affine connections (for invariant connections see also e.g. Helgason~\cite{ref6}). The obtained invariant metrizability equations are mere applications of the formulas for the corresponding Lie derivatives. 

\subsection{The Lie derivative of $(0,2)$-tensor fields}

Consider a $(0,2)$-tensor field $g$ on $X$ and a vector field $\xi $ on $X$, and denote by ${\partial _\xi }g$ the Lie derivative of $g$ by $\xi $. If $g$ and $\xi $ are expressed in a chart $(U,\varphi )$, $\varphi  = ({x^i})$, by 
\begin{eqnarray}
g = {g_{ij}}d{x^i} \otimes d{x^j},\quad \xi  = {\xi ^i}\frac{\partial }{{\partial {x^i}}}, \label{eq5}
\end{eqnarray}
then ${\partial _\xi }g$ has the well-known expression 
\begin{eqnarray}
{\partial _\xi }g = \left( {\frac{{\partial {g_{kl}}}}{{\partial {x^p}}}{\xi ^p} + {g_{il}}\frac{{\partial {\xi ^i}}}{{\partial {x^k}}} + {g_{kj}}\frac{{\partial {\xi ^j}}}{{\partial {x^l}}}} \right)d{x^k} \otimes d{x^l}. \label{eq6}
\end{eqnarray}

\subsection{The Lie derivative of affine connections}

Let $\mathcal{V}X$ denote the module of vector fields on $X$. Recall that an {\it affine connection} on $X$ is a mapping $\mathcal{V}X \times \mathcal{V}X \ni (\xi ,\zeta ) \to \nabla (\xi ,\zeta ) = {\nabla _\xi }(\zeta ) \in \mathcal{V}X$, such that
\begin{enumerate}
\item ${\nabla _{f\xi  + g\zeta }} = f{\nabla _\xi } + g{\nabla _\zeta }$ , 
\item ${\nabla _\lambda }(\xi  + \zeta ) = {\nabla _\lambda }(\xi ) + {\nabla _\lambda }(\zeta )$, 
\item ${\nabla _\xi }(f\zeta ) = f{\nabla _\xi }(\zeta ) + \xi (f)\zeta $ 
\end{enumerate}
for all functions $f, g$ on $X$ and all vector fields $\lambda ,\xi ,\zeta  \in \mathcal{V}X$. $\nabla $ is said to be {\it symmetric}, if ${\nabla _\xi }(\zeta ) - {\nabla _\zeta }(\xi ) = [\xi ,\zeta ]$ for all $\xi $ and $\zeta $. 

	Given a chart $(U,\varphi )$, $\varphi  = ({x^i})$, then the local coordinate expression will be
\begin{eqnarray}
{\nabla _\xi }(\zeta ) = {\xi ^k}\left( {\frac{{\partial {\zeta ^i}}}{{\partial {x^k}}} + \nabla _{kl}^i{\zeta ^l}} \right)\frac{\partial }{{\partial {x^i}}}\label{eq7}
\end{eqnarray}
with
\begin{eqnarray}
{\nabla _{\partial /\partial {x^j}}}\left( {\frac{\partial }{{\partial {x^k}}}} \right) = \nabla _{jk}^i\frac{\partial }{{\partial {x^i}}},\quad \xi  = {\xi ^k}\frac{\partial }{{\partial {x^k}}},\quad \zeta  = {\zeta ^k}\frac{\partial }{{\partial {x^k}}}.\label{eq8}
\end{eqnarray}
The mapping ${\nabla _\xi }:\mathcal{V}X \to \mathcal{V}X$ is called a {\it covariant derivative} of a vector field with respect to the vector field $\xi $.

Let $\alpha :X \to X$ be a diffeomorphism. For every vector field $\xi $ we define a vector field $\xi {}_{(\alpha )}$ on X, the {\it pushforward} of $\xi $ by $\alpha $, by $\xi {}_{(\alpha )} = T\alpha  \cdot (\xi  \circ {\alpha ^{ - 1}})$. More completely, the value of $\xi {}_{(\alpha )}$ at a point $x \in X$ is given by 
\begin{eqnarray}
\xi {}_{(\alpha )}(x) = {T_{{\alpha ^{ - 1}}(x)}}\alpha  \cdot \xi ({\alpha ^{ - 1}}(x)). \label{eq9}
\end{eqnarray}
If $(U,\varphi )$, $\varphi  = ({x^i})$, and $(V,\psi )$, $\psi  = ({y^i})$, are two charts on $X$ such that $\alpha (V) = U$, and $x \in U$ is a point, and if $\xi ({\alpha ^{ - 1}}(x))$ is expressed by 
\begin{eqnarray}
\xi ({\alpha ^{ - 1}}(x)) = {\xi ^k}({\alpha ^{ - 1}}(x)){\left( {\frac{\partial }{{\partial {y^k}}}} \right)_{{\alpha ^{ - 1}}(x)}}, \label{eq10}
\end{eqnarray}
then from (\ref{eq9}) 
\begin{eqnarray}
\xi {}_{(\alpha )}(x) = {\left( {\frac{{\partial ({x^k}\alpha {\psi ^{ - 1}})}}{{\partial {y^l}}}} \right)_{\psi {\alpha ^{ - 1}}(x)}}{\xi ^l}({\alpha ^{ - 1}}(x)){\left( {\frac{\partial }{{\partial {x^k}}}} \right)_x}.\label{eq11}
\end{eqnarray}

Let $\nabla $ be an affine connection on $X$. Every diffeomorphism $\alpha :X \to X$ defines a mapping ${\nabla ^{(\alpha )}}:\mathcal{V}X \times \mathcal{V}X \to \mathcal{V}X$ by ${\nabla ^{(\alpha )}}(\xi ,\zeta ) = \nabla {({\xi _{(\alpha )}},{\zeta _{(\alpha )}})_{({\alpha ^{ - 1}})}}$. Denoting $\nabla _\xi ^{(\alpha )}(\zeta ) = {\nabla ^{(\alpha )}}(\xi ,\zeta )$, we can equivalently write 
\begin{eqnarray}
\nabla _\xi ^{(\alpha )}(\zeta )(y) = {({\nabla _{{\xi _{(\alpha )}}}}({\zeta _{(\alpha )}}))_{({\alpha ^{ - 1}})}}(y) = {T_{\alpha (y)}}{\alpha ^{ - 1}} \cdot {\nabla _{{\xi _{(\alpha )}}}}({\zeta _{(\alpha )}})(\alpha (y))\label{eq12}
\end{eqnarray}
for every point $y \in X$, and all $\xi $, $\zeta $. If $y = {\alpha ^{ - 1}}(x)$, then 
\begin{eqnarray}
\nabla _\xi ^{(\alpha )}(\zeta )({\alpha ^{ - 1}}(x)) = {({\nabla _{{\xi _{(\alpha )}}}}({\zeta _{(\alpha )}}))_{({\alpha ^{ - 1}})}}({\alpha ^{ - 1}}(x)) = {T_x}{\alpha ^{ - 1}} \cdot {\nabla _{{\xi _{(\alpha )}}}}({\zeta _{(\alpha )}})(x). \label{eq13}
\end{eqnarray}
From definition, it is straightforward to see that ${\nabla ^{(\alpha )}}$ again will be an affine connection, which is said to be {\it associated} with $\nabla $ by $\alpha $.

If $\nabla _{jk}^i$ are the components of $\nabla $ on $U$, then the new affine connection will be given on $V$ by
\begin{eqnarray}
{\nabla ^{(\alpha )}}\left( {\xi ,\zeta } \right) = {\xi ^i}\left( {\frac{{\partial {\zeta ^k}}}{{\partial {x^i}}} + {\zeta ^j}\nabla _{ij}^{(\alpha )k}} \right)\frac{\partial }{{\partial {x^k}}},\label{eq14}
\end{eqnarray}
where 
\begin{eqnarray}
&&\hspace{-0.5cm}\nabla _{ij}^{(\alpha )k}(y) = {\left( {\frac{{\partial ({y^k}{\alpha ^{ - 1}}{\varphi ^{ - 1}})}}{{\partial {x^s}}}} \right)_{\varphi (\alpha (y))}} \nonumber \\ 
&&\hspace{-0.5cm}\left( {{{\left( {\frac{{\partial ({x^a}\alpha {\psi ^{ - 1}})}}{{\partial {y^i}}}} \right)}_{\psi (y)}}{{\left( {\frac{{\partial ({x^b}\alpha {\psi ^{ - 1}})}}{{\partial {y^j}}}} \right)}_{\psi (y)}}{\nabla ^s}_{ab}\left( {\alpha \left( y \right)} \right) + {{\left( {\frac{{{\partial ^2}({x^s}\alpha {\psi ^{ - 1}})}}{{\partial {y^i}\partial {y^j}}}} \right)}_{\psi (y)}}} \right).   \label{eq15}
\end{eqnarray}

In particular, if $\nabla $ is symmetric, ${\nabla ^{(\alpha )}}$ is also symmetric. If $\alpha  = {\operatorname{id} _X}$, (\ref{eq14}) reduces to the transformation formula for components of an affine connection. 

One can easily show that if $\nabla $ is metrizable and $\nabla  = {}^g\Gamma $ for a metric field $g$, then ${\nabla ^{(\alpha )}}$ is also metrizable and ${\nabla ^{(\alpha )}} = {}^{\alpha *g}\Gamma $. 

We now study transformation properties of affine connections with respect to one-parameter transformation groups. We start with the invariant expression (\ref{eq6}), considered at a fixed point $x$ belonging to the domain of definition $U$ of a chart $(U,\varphi )$, $\varphi  = ({x^i})$. 
If $\lambda $ is a vector field on $X$ and ${\alpha _t}$ is the one-parameter group of $\lambda $, 
then formula (\ref{eq14}) applies to the associated connection $\nabla {}^{({\alpha _t})}$, 
defined on a neighborhood of $x$ for all sufficiently small $t$. 
For all $\xi $ and $\zeta $ we have a curve $t \to \nabla _\xi ^{({\alpha _t})}(\zeta )(x)$ in the tangent space ${T_x}X$. We define the {\it Lie derivative} ${\partial _\lambda }\nabla $ of $\nabla $ by $\lambda $ by 
\begin{eqnarray}
({\partial _\lambda }\nabla )(\xi ,\zeta )(x) = {\left( {\frac{d}{{dt}}\nabla _\xi ^{({\alpha _t})}(\zeta )(x)} \right)_0}. \label{eq16}
\end{eqnarray}
In this formula 
\begin{eqnarray}
\nabla _\xi ^{({\alpha _t})}(\zeta )(x) = {\xi ^k}(x)\left( {{{\left( {\frac{{\partial {\zeta ^i}}}{{\partial {x^k}}}} \right)}_{\varphi (x)}} 
+ {\nabla^{({\alpha_t})_i}}_{kl}(x){\zeta ^l}(x)} \right)
{\left( {\frac{\partial }{{\partial {x^i}}}} \right)_x}, \label{eq17}
\end{eqnarray}
and by (\ref{eq15}) 
\begin{eqnarray}
&&\hspace{-0.5cm}{\nabla ^{({\alpha _t})}}_{jk}^i\;\,(x) = {\left( {\frac{{\partial ({x^i}{\alpha _{ - t}}{\varphi ^{ - 1}})}}{{\partial {x^s}}}} \right)_{\varphi ({\alpha _t}(x))}} \nonumber \\
&&\hspace{-0.5cm}\cdot \left( {{{\left( {\frac{{\partial ({x^l}{\alpha _t}{\varphi ^{ - 1}})}}{{\partial {x^j}}}} \right)}_{\varphi (x)}}{{\left( {\frac{{\partial ({x^t}{\alpha _t}{\varphi ^{ - 1}})}}{{\partial {x^k}}}} \right)}_{\varphi (x)}}{\nabla ^{({\alpha _t})}}_{lt}^s({\alpha _t}(x))} \right. + \left. {{{\left( {\frac{{{\partial ^2}({x^s}{\alpha _t}{\varphi ^{ - 1}})}}{{\partial {x^j}\partial {x^k}}}} \right)}_{\varphi (x)}}} \right). \label{eq18}  
\end{eqnarray}
Writing 
\begin{eqnarray}
\lambda  = {\lambda ^k}\frac{\partial }{{\partial {x^i}}}, \label{eq19}
\end{eqnarray}
and differentiating (\ref{eq18}) with respect to $t$ at $t = 0$ we get the following expression for the Lie derivative ${\partial _\lambda }\nabla $
\begin{eqnarray}
{({\partial _\lambda }\nabla )_\xi }(\zeta ) = \left( { - \frac{{\partial {\lambda ^i}}}{{\partial {x^s}}}\nabla _{jk}^s + \frac{{\partial {\lambda ^s}}}{{\partial {x^j}}}\nabla _{sk}^i + \frac{{\partial {\lambda ^m}}}{{\partial {x^k}}}\nabla _{jm}^i + \frac{{\partial \nabla _{jk}^i}}{{\partial {x^q}}}{\lambda ^q} + \frac{{{\partial ^2}{\lambda ^i}}}{{\partial {x^j}\partial {x^k}}}} \right){\xi ^k}{\zeta ^l}{\left( {\frac{\partial }{{\partial {x^i}}}} \right)_x}. \label{eq20}
\end{eqnarray} 

Formula (\ref{eq20}) shows that ${\partial _\lambda }\nabla $ is a tensor field of type $(1,2)$. 

We end this section with a global formula on the structure of the Lie derivative ${\partial _\lambda }\nabla $, and an elementary property of connections, associated with diffeomorphisms.

\begin{lemma}
For any vector fields $\xi ,\zeta ,\lambda$ on $X$
\begin{eqnarray}
({\partial _\xi }\nabla )(\zeta ,\lambda ) = {\partial _\xi }(\nabla (\zeta ,\lambda )) - \nabla ({\partial _\xi }\zeta ,\lambda ) - \nabla (\zeta ,{\partial _\xi }\lambda ). \label{eq21}
\end{eqnarray}
\end{lemma}

\begin{lemma}
  If $\nabla $ is metrizable and $\nabla  = {}^g\Gamma $ for a metric field $g$, then ${\nabla ^{(\alpha )}}$ is also metrizable and ${\nabla ^{(\alpha )}} = {}^{\alpha *g}\Gamma $. 
\end{lemma}

\section{Invariant metrizability problem}
Recall that a $(0,2)$-tensor field $g$ on $X$ is {\it invariant} with respect to a diffeomorphism $\alpha $ of $X$, if its pull-back $\alpha *g$ satisfies 
\begin{eqnarray}
\alpha *g = g. \label{eq22}	
\end{eqnarray}
Similarly an affine connection $\nabla $ on X is {\it invariant} with respect to $\alpha $, if 
\begin{eqnarray}
\nabla ^\alpha  = \nabla . \label{eq23}
\end{eqnarray}
The following is an immediate consequence of definitions. 

\begin{lemma} \label{lem3}
For any diffeomorphism $\alpha $of $X$, the Levi-Civita connection ${}^g\Gamma$ satisfies 
\begin{eqnarray}
{{(^g}\Gamma )^{(\alpha )}} = {\;^{\alpha *g}}\Gamma .\label{eq24}
\end{eqnarray}
In particular, if $g$ is $G$-invariant, then ${}^g\Gamma $ is also $G$-invariant. 
\end{lemma}

We have already seen that the definitions of $G$-invariance extend naturally to invariance with respect to one-parameter groups of vector fields. In the following lemma we combine invariance and metrizability conditions and get, in our standard notation, a system of partial differential equations, characterizing the $\xi $-metrizability problem in terms of charts. 

\begin{lemma}  ($\xi $-metrizability equations)  \label{lem4}
Let $(U,\varphi )$, $\varphi  = ({x^i})$, be a chart on X, and let $\xi $ be a vector field, expressed by 
\begin{eqnarray}
\xi  = {\xi ^k}\frac{\partial }{{\partial {x^i}}}. \label{eq25}
\end{eqnarray}
Then the $\xi $-metrizability problem is characterized in terms of $(U,\varphi )$ by the following equations: \\ 
1. Equation for $\xi $-invariant affine connections 
\begin{eqnarray}
- \frac{{\partial {\xi ^i}}}{{\partial {x^m}}}\nabla _{jk}^m + \frac{{\partial {\xi ^m}}}{{\partial {x^j}}}\nabla _{mk}^i + \frac{{\partial {\xi ^m}}}{{\partial {x^k}}}\nabla _{jm}^i + \frac{{\partial \nabla _{jk}^i}}{{\partial {x^m}}}{\xi ^m} + \frac{{{\partial ^2}{\xi ^i}}}{{\partial {x^j}\partial {x^k}}} = 0.	\label{eq26}
\end{eqnarray}
2. Equation for $\xi $-invariant (0,2)-tensor fields 
\begin{eqnarray}
\frac{{\partial {g_{ij}}}}{{\partial {x^m}}}{\xi ^m} + {g_{im}}\frac{{\partial {\xi ^m}}}{{\partial {x^j}}} + {g_{mj}}\frac{{\partial {\xi ^m}}}{{\partial {x^i}}} = 0. \label{eq27}
\end{eqnarray}
3. The metrizability equation 
\begin{eqnarray}
\frac{{\partial {g_{ij}}}}{{\partial {x^k}}} = {g_{im}}\nabla _{jk}^m + {g_{jm}}\nabla _{ik}^m,\quad {g_{ij}} = {g_{ji}},\quad \det {g_{ij}} \ne 0.		\label{eq28}
\end{eqnarray}
\end{lemma}

Clearly, Lemma \ref{lem4} also describes {\it G-metrizability} whenever $G$ is a {\it connected} Lie group; in this case we take for $\xi $ the {\it generators} of the group action of the Lie group $G$ on $X$. 

\section{$\operatorname{SO} (3)$-Metrizability: Example 3D}
In this section we consider the open set $X = {R^3}\backslash \left\{ {(0,0,0)} \right\}$ in the Euclidean space ${R^3}$ with its canonical manifold structure and the canonical left action of the rotation group $\operatorname{SO} (3)$. 
Given an $\operatorname{SO} (3)$-invariant affine connection $\nabla $ on $X$, we find restrictions to $\nabla$ (integrability conditions) and all $\operatorname{SO} (3)$-invariant metric fields $g$ whose Levi-Civita connection ${}^g\Gamma $ coincides with $\nabla$. 
\subsection{Spherical atlas}
We introduce two spherical charts on ${R^3}$, defining a smooth atlas on $X$. We need these charts to simplify further calculations and also for some elementary global constructions. 

Consider the mapping ${R^3} \ni (r,\varphi ,\vartheta ) \to (x(r,\varphi ,\vartheta ),y(r,\varphi ,\vartheta ),z(r,\varphi ,\vartheta )) \in {R^3}$ defined by the equations 
\begin{eqnarray}
x = r\sin \vartheta \cos \varphi ,\quad y = r\sin \vartheta \sin \varphi ,\quad z = r\cos \vartheta .		\label{eq29}
\end{eqnarray}
Since the Jacobi determinant of this mapping is $ - {r^2}\sin \vartheta $, 
equations (\ref{eq29}) define a local diffeomorphism at every point of the open set in ${R^3}$ where $r \ne 0$ and $\sin \vartheta  \ne 0$. 

It can be easily verified that the restriction of the mapping (\ref{eq29}) to the open set $\mathcal{U} = (0,\infty ) \times (0,2\pi ) \times (0,\pi )$ is a diffeomorphism of $\mathcal{U}$ and the subset $U$ of ${R^3}$ defined as \\
$U = {R^3}\backslash \left\{ {(x,y,z) \in {R^3}|x \geqslant 0,y = 0} \right\}$. 
The inverse diffeomorphism is the mapping $c$, where 
\begin{eqnarray}
r = \sqrt {{x^2} + {y^2} + {z^2}} , \nonumber \\
  \varphi  = \left\{ {\,\begin{array}{*{20}{c}}
  {\arccos \frac{x}{{\sqrt {{x^2} + {y^2}} }},\quad x < 0,} \nonumber\\ 
  {\arcsin \frac{y}{{\sqrt {{x^2} + {y^2}} }}\quad y > 0,\,\;}\nonumber \\ 
  {\arcsin \frac{y}{{\sqrt {{x^2} + {y^2}} }}\quad y < 0,\,\;} 
\end{array}} \right. \nonumber \\
  \vartheta  = \arccos \frac{z}{{\sqrt {{x^2} + {y^2} + {z^2}} }}.  \label{eq30} 
\end{eqnarray}

This construction can be modified by means of a rotation $\nu $ of ${R^3}$, expressed by the equations $x \circ \nu  =  - x$, $y \circ \nu  =  - z$, and $z \circ \nu  =  - y$.
We define $\bar U = {\nu ^{ - 1}}(U) = {R^3}\backslash \left\{ {(x,y,z) \in {R^3}|x \leqslant 0,z = 0} \right\}$ 
and $\bar \Phi  = \Phi  \circ \nu  = (\bar r,\bar \varphi ,\bar \vartheta )$, 
where $\bar r = r \circ \nu $, $\bar \varphi  = \varphi  \circ \nu $, 
$\bar \vartheta  = \vartheta  \circ \nu$; then 
\begin{eqnarray}
  \bar r = \sqrt {{x^2} + {y^2} + {z^2}} , \nonumber\\
  \bar \varphi  = \left\{ {\,\begin{array}{*{20}{c}}
  {\arccos \left( { - \frac{x}{{\sqrt {{x^2} + {z^2}} }}} \right),\quad x > 0,} \nonumber\\ 
  {\arcsin \left( { - \frac{z}{{\sqrt {{x^2} + {z^2}} }}} \right)\quad z < 0,\,\;} \nonumber\\ 
  {\arcsin \left( { - \frac{z}{{\sqrt {{x^2} + {z^2}} }}} \right)\quad z > 0,\,\;} 
\end{array}} \right.  \nonumber\\
  \bar \vartheta  = \arccos \left( { - \frac{y}{{\sqrt {{x^2} + {y^2} + {z^2}} }}} \right).  \label{eq31}
\end{eqnarray}
The inverse transformation of (\ref{eq31}) can be easily determined by replacing  $x \to  - x$, $y \to  - z$, $z \to  - y$ in (\ref{eq29}). 
We get 
\begin{eqnarray}
x =  - \bar r\sin \bar \vartheta \cos \bar \varphi ,\quad z =  - \bar r\sin \bar \vartheta \sin \bar \varphi ,\quad y =  - \bar r\cos \bar \vartheta .		\label{eq32}
\end{eqnarray}
The pairs $(U,\Phi )$, $\Phi  = (r,\varphi ,\vartheta )$, 
and $(\bar U,\bar \Phi )$, $\bar \Phi  = (\bar r,\bar \varphi ,\bar \vartheta )$, 
are charts on $X$, called the {\it first spherical chart}, and the {\it second spherical chart}, respectively. 
Clearly, $\bar \Phi (\bar U) = \Phi (U) = \mathcal{U}$. 
These two charts form a smooth atlas on $X$, called the {\it spherical atlas}. 
The coordinate transformation $\bar \Phi {\Phi ^{ - 1}}$ is expressed by 
\begin{eqnarray}
\bar r = r,\quad \sin \bar \varphi  =  - \frac{{\cos \vartheta }}{{\sqrt {1 - {{\sin }^2}\vartheta {{\sin }^2}\varphi } }},\quad \cos \bar \vartheta  =  - \sin \vartheta \sin \varphi .		\label{eq33}
\end{eqnarray} 
The following formulas, related to spherical charts, are given here for the reference. 

\begin{lemma} \label{lem5}
At every point $(x,y,z) \in U \cap \bar U$
\begin{eqnarray}
d\vartheta  \otimes d\vartheta  + {\sin ^2}\vartheta d\varphi  \otimes d\varphi  = d\bar \vartheta  \otimes d\bar \vartheta  + {\sin ^2}\bar \vartheta d\bar \varphi  \otimes d\bar \varphi .		\label{eq34}
\end{eqnarray}
\end{lemma}
The generators of the rotations in ${R^3}$ around coordinate axes are expressed in the Cartesian coordinates by 
\begin{eqnarray}
\xi  = x\frac{\partial }{{\partial y}} - y\frac{\partial }{{\partial x}},\quad \zeta  = y\frac{\partial }{{\partial z}} - z\frac{\partial }{{\partial y}},\quad \lambda  = z\frac{\partial }{{\partial x}} - x\frac{\partial }{{\partial z}}.	\label{eq35}
\end{eqnarray}

\begin{lemma} \label{lem6}
 In the first spherical coordinates 
\begin{eqnarray}
\xi  = \frac{\partial }{{\partial \varphi }},\quad \zeta  =  - \sin \varphi \frac{\partial }{{\partial \vartheta }} - \cot \vartheta \cos \varphi \frac{\partial }{{\partial \varphi }},\quad \lambda  = \cos \varphi \frac{\partial }{{\partial \vartheta }} - \cot \vartheta \sin \varphi \frac{\partial }{{\partial \varphi }}.	\label{eq36}
\end{eqnarray}
\end{lemma}

\subsection{$\operatorname{SO} (3)$-invariant $(0,2)$-tensor fields}  
Let $g$ be a $(0,2)$-tensor field $g$ on the manifold $X$, given in the first spherical chart by 
\begin{eqnarray}
&&g = {g_{rr}}dr \otimes dr + {g_{r\varphi }}dr \otimes d\varphi  + {g_{r\vartheta }}dr \otimes d\vartheta  + {g_{\varphi r}}d\varphi  \otimes dr + {g_{\varphi \varphi }}d\varphi  \otimes d\varphi  \nonumber  \\
&&\quad  + {g_{\varphi \vartheta }}d\varphi  \otimes d\vartheta  + {g_{\vartheta r}}d\vartheta  \otimes dr + {g_{\vartheta \varphi }}d\vartheta  \otimes d\varphi  + {g_{\vartheta \vartheta }}d\vartheta  \otimes d\vartheta .  \label{eq37} 
\end{eqnarray}
The following lemma describes solutions ${g_{rr}},{g_{r\varphi }},{g_{r\vartheta }},{g_{\varphi \varphi }},{g_{\varphi \vartheta }},{g_{\vartheta \vartheta }}$ of the Killing equations ${\partial _\xi }g = 0$, ${\partial _\zeta }g = 0$, and ${\partial _\lambda }g = 0$ for the generators (\ref{eq36}). 

\begin{lemma} \label{lem7}
If a $(0,2)$-tensor field $g$ on $X$ is invariant with respect to rotations, then in the first spherical coordinates
\begin{eqnarray} 
g = P(r)dr \otimes dr + Q(r)(d\vartheta  \otimes d\vartheta  + {\sin ^2}\vartheta d\varphi  \otimes d\varphi ), \label{eq38}
\end{eqnarray}
where $P$ and $Q$ are functions, depending on $r$ only. 
\end{lemma}
Using the spherical atlas on $X$ and (\ref{eq34}) we can globalize formula (\ref{eq38}) as follows. 

\begin{lemma} \label{lem8}
Let 
\begin{eqnarray}
{g_U} = P(r)dr \otimes dr + Q(r)(d\vartheta  \otimes d\vartheta  + {\sin ^2}\vartheta d\varphi  \otimes d\varphi )			\label{eq39}
\end{eqnarray}
be an $\operatorname{SO} (3)$-invariant $(0,2)$-tensor field on $U$, and let
\begin{eqnarray} 
{\bar g_{\bar U}} = \bar P(\bar r)d\bar r \otimes d\bar r + \bar Q(\bar r)(d\bar \vartheta  \otimes d\bar \vartheta  + {\sin ^2}\bar \vartheta d\bar \varphi  \otimes d\bar \varphi )	 	\label{eq40}
\end{eqnarray}
be an $\operatorname{SO} (3)$-invariant $(0,2)$-tensor field on $\bar U$. The following two conditions are equivalent:\\ 
1.	${g_U} = {\bar g_{\bar U}}$ on $U \cap \bar U$. \\
2.	$P = \bar P$ and $Q = \bar Q$ on $(0,\infty )$. \\
\end{lemma} 
Lemma \ref{lem8} constitutes a one-to-one correspondence between $\operatorname{SO} (3)$-invariant $(0,2)$-tensor fields on $X$ and the pairs of everywhere non-zero functions $(P,Q)$, defined on the set of positive real numbers $(0,\infty )$. 
In particular, Lemma \ref{lem8} proves the following theorem. 

\begin{theorem} \label{th9}
The manifold $X$, endowed with any $\operatorname{SO} (3)$-invariant metric field, is a warped product of the manifolds $(0,\infty )$ and ${S^2}$, considered with their canonical metric fields.  
\end{theorem}

\subsection{$\operatorname{SO} (3)$-invariant affine connections}
To characterize $\operatorname{SO} (3)$-invariant affine connections on $X$, 
we apply invariance conditions (\ref{eq26}). 
Considering separately rotations around $x$-axis, $y$-axis and $z$-axis and the corresponding invariance equations, defined by generators (\ref{eq36}), we can prove the following result. 

\begin{lemma}\label{lem10}
Every $\operatorname{SO}(3)$-invariant affine connection $\nabla $ on ${R^3}$ has the components 
\begin{eqnarray}
&&\nabla _{11}^1 = A_{11}^1(r),\quad \nabla _{12}^1 = 0,\quad \nabla _{13}^1 = 0,\quad \nabla _{22}^1 = A_{22}^1(r),\quad \nabla _{23}^1 = 0, \nonumber \\
&&\quad \nabla _{33}^1(r,\vartheta ) = {\sin ^2}\vartheta  \cdot A_{22}^1(r), \nonumber \\
&&\nabla _{11}^2 = 0,\quad \nabla _{12}^2 = A_{12}^2(r),\quad \nabla _{13}^2(r,\vartheta ) = A(r)\sin \vartheta , \nonumber \\
&&\quad \nabla _{22}^2 = 0,\quad \nabla _{23}^2 = 0,\quad \nabla _{33}^2(\vartheta ) =  - \sin \vartheta \cos \vartheta , \nonumber \\
&&\nabla _{11}^3 = 0,\quad \nabla _{12}^3(r,\vartheta ) =  - \frac{{A(r)}}{{\sin \vartheta }},\quad \nabla _{13}^3 = A_{12}^2(r),\quad \nabla _{22}^3 = 0, \nonumber \\
&&\quad \nabla _{23}^3(r,\vartheta ) = \cot \vartheta ,\quad \nabla _{33}^3 = 0,   \label{eq41}
\end{eqnarray}
where $A_{11}^1$, $A_{22}^1$, $A_{12}^2$, and $A$ are arbitrary functions of the variable $r$. 
\end{lemma}

\subsection{$\operatorname{SO} (3)$-metrizability } 
In the following two theorems we give the solution to the $\operatorname{SO} (3)$-metrizability problem. The proof can be given by direct analysis of the $\operatorname{SO} (3)$-metrizability conditions. 

\begin{theorem} \label{th11} 
Let $\nabla $ be an affine connection. The following two conditions are equivalent:\\
1. $\nabla $ is $\operatorname{SO} (3)$-invariant and $\operatorname{SO} (3)$-metrizable. \\
2.	The components of $\nabla $ satisfy
\begin{eqnarray}
&&\nabla _{11}^1 = \nabla _{11}^1(r),\quad \nabla _{12}^1 = 0,\quad \nabla _{13}^1 = 0,\quad \nabla _{22}^1 =  - \frac{L}{K}A_{12}^2{\text{exp}}\left( {2\int_1^r {(A_{12}^2(t) - A_{11}^1(t))dt} } \right), \nonumber \\
&&\quad \nabla _{23}^1 = 0,\quad \nabla _{33}^1 =  - \frac{L}{K}{\sin ^2}\vartheta  \cdot A_{12}^2{\text{exp}}\left( {2\int_1^r {A_{12}^2(t) - A_{11}^1(t)dt} } \right), \nonumber \\
&&\nabla _{11}^2 = 0,\quad \nabla _{12}^2 = A_{12}^2(r),\quad \nabla _{13}^2 = 0,\quad \nabla _{22}^2 = 0,\quad \nabla _{23}^2 = 0,\quad \nabla _{33}^2 =  - \sin \vartheta \cos \vartheta , \nonumber \\
&&\quad \nabla _{11}^3 = 0,\quad \nabla _{12}^3 = 0,\quad \nabla _{13}^3 = \nabla _{12}^2(r),\quad \nabla _{22}^3 = 0,\quad \nabla _{23}^3 = \cot \vartheta ,\quad \nabla _{33}^3 = 0  \label{eq42} 
\end{eqnarray}
for some nonzero constants $K,L \in R$. 
\end{theorem}

Conditions (\ref{eq42}) are $\operatorname{SO} (3)$-{\it metrizability conditions} for $\nabla $. 
It also follows from (\ref{eq42}) that every $\operatorname{SO} (3)$-metrizable affine connection depends on two arbitrary functions $A_{11}^1 = A_{11}^1(r)$ and $A_{12}^2 = A_{12}^2(r)$ only. 

\begin{theorem} \label{th12}  
If an $\operatorname{SO} (3)$-invariant affine connection $\nabla $ satisfies the local metrizability conditions (\ref{eq42}), then the $\operatorname{SO} (3)$-metrizability problem has a solution 
\begin{eqnarray}
g = P(r)dr \otimes dr + Q(r)(d\vartheta  \otimes d\vartheta  + {\sin ^2}\vartheta d\varphi  \otimes d\varphi ),		\label{eq43}
\end{eqnarray}
where 
\begin{eqnarray}
P(r) = K{\text{exp}}\left( {2\int_1^r {A_{11}^1(t)dt} } \right),\quad Q(r) = L{\text{exp}}\left( {2\int_1^r {A_{12}^2(t)dt} } \right).		\label{eq44}
\end{eqnarray}
\end{theorem}

Formula (\ref{eq44}) describes {\it globally defined} metric fields on $X$ (cf. Lemma \ref{lem8}). 
Note that $\operatorname{SO} (3)$-metrizability does not influence the signature of $g$.

\section{$\operatorname{SO} (3)$-Metrizability: Example 4D}
In this section $X = R \times ({R^3}\backslash \left\{ {(0,0,0)} \right\})$, and we consider this open subset of ${R^4}$ with its standard manifold structure and the canonical left action of the rotation group $\operatorname{SO} (3)$ on the second factor. $X$ is homeomorphic with ${R^2} \times {S^2}$, where ${S^2}$ is the 2-dimensional unit sphere. A homeomorphism can be constructed from the homeomorphism 
\begin{eqnarray}
R \times ({R^3}\backslash \left\{ {(0,0,0)} \right\}) \ni (t,x,y,z) \to (t,\theta (x,y,z)) \in R \times (0,\infty ) \times {S^2},	\label{eq45}
\end{eqnarray}
which is defined by the homeomorphism 
\begin{eqnarray}
{R^3}\backslash \left\{ {(0,0,0)} \right\} \ni (x,y,z) \to \theta (x,y,z) = \left( {r,\left( {\frac{x}{r},\frac{y}{r},\frac{z}{r}} \right)} \right) \in (0,\infty ) \times {S^2},	\label{eq46}
\end{eqnarray}
where $r = \sqrt {{x^2} + {y^2} + {z^2}} $. We find all $\operatorname{SO} (3)$-invariant affine connections $\nabla $ on $X$, and all $\operatorname{SO} (3)$-metrizable metric fields on $X$. 

\subsection{$\operatorname{SO} (3)$-invariant $(0,2)$-tensor fields}
Consider $(0,2)$-tensor fields g on the manifold $X$, given in the first spherical chart by 
\begin{eqnarray}
&&g = {g_{tt}}dt \otimes dt + {g_{tr}}dt \otimes dr + {g_{t\varphi }}dt \otimes d\varphi  + {g_{t\vartheta }}dt \otimes d\vartheta  \nonumber \\
&&\quad  + {g_{rt}}dr \otimes dt + {g_{rr}}dr \otimes dr + {g_{r\varphi }}dr \otimes d\varphi  + {g_{r\vartheta }}dr \otimes d\vartheta  \nonumber \\
&&\quad  + {g_{\varphi t}}d\varphi  \otimes dt + {g_{\varphi r}}d\varphi  \otimes dr + {g_{\varphi \varphi }}d\varphi  \otimes d\varphi  + {g_{\varphi \vartheta }}d\varphi  \otimes d\vartheta  \nonumber \\
&&\quad  + {g_{\vartheta t}}d\vartheta  \otimes dt + {g_{\vartheta r}}d\vartheta  \otimes dr + {g_{\vartheta \varphi }}d\vartheta  \otimes d\varphi  + {g_{\vartheta \vartheta }}d\vartheta  \otimes d\vartheta . \label{eq47} 
\end{eqnarray}
The following lemma describes solutions ${g_{tt}},{g_{tr}},{g_{t\varphi }},{g_{t\vartheta }},{g_{rr}},{g_{r\varphi }},{g_{r\vartheta }},{g_{\varphi \varphi }},$${g_{\varphi \vartheta }},{g_{\vartheta \vartheta }}$of the Killing equations ${\partial _\xi }g = 0$, ${\partial _\zeta }g = 0$, and ${\partial _\lambda }g = 0$, where $\xi $, $\zeta $ and $\lambda $ are the generators of rotations in $X$ (Lemma \ref{lem6}). 

\begin{lemma}\label{lem13}
If a $(0,2)$-tensor field $g$ on $X$ is invariant with respect to rotations, then in the first spherical coordinates 
\begin{eqnarray}
&&g = {g_{tt}}(t,r)dt \otimes dt + {g_{tr}}(t,r)(dt \otimes dr + dr \otimes dt) \nonumber \\
&&\quad  + {g_{rr}}(t,r)dr \otimes dr + Q(t,r)(d\vartheta  \otimes d\vartheta  + {\sin ^2}\vartheta d\varphi  \otimes d\varphi ).  \label{eq48} 
\end{eqnarray}
\end{lemma}
Note that $\det {g_{ij}} = ({g_{tt}}{g_{rr}} - g_{tr}^2){Q^2}{\sin ^2}\vartheta $, thus the non-singularity of the tensor field $g$ is equivalent with the conditions ${g_{tt}}{g_{rr}} - g_{tr}^2 \ne 0$, $Q \ne 0$. The first of these conditions implies that at least one of the components ${g_{tt}},{g_{rr}},{g_{tr}}$ must always be different from 0. 

Using the spherical atlas on $X$ we can easily globalize local expression (\ref{eq48}) as follows. 

\begin{lemma}\label{lem14}  Let 
\begin{eqnarray}
&&{g_U} = {g_{tt}}(t,r)dt \otimes dt + {g_{tr}}(t,r)(dt \otimes dr + dr \otimes dt)\nonumber  \\
&&\quad  + {g_{rr}}(t,r)dr \otimes dr + Q(t,r)(d\vartheta  \otimes d\vartheta  + {\sin ^2}\vartheta d\varphi  \otimes d\varphi ) \label{eq49}
\end{eqnarray}
be an $\operatorname{SO} (3)$-invariant $(0,2)$-tensor field on $U$, and let 
\begin{eqnarray}
&&{{\bar g}_{\bar U}} = {g_{\bar t\bar t}}(\bar t,\bar r)d\bar t \otimes d\bar t + {g_{\bar t\bar r}}(\bar t,\bar r)(d\bar t \otimes d\bar r + d\bar r \otimes d\bar t)  \nonumber \\
&&\quad  + {g_{\bar r\bar r}}(\bar t,\bar r)d\bar r \otimes d\bar r + Q(\bar t,\bar r)(d\bar \vartheta  \otimes d\bar \vartheta  + {\sin ^2}\bar \vartheta d\bar \varphi  \otimes d\bar \varphi ) \label{eq50}
\end{eqnarray}
be an $\operatorname{SO} (3)$-invariant $(0,2)$-tensor field on $\bar U$. The following two conditions are equivalent:\\ 
1.	${g_U} = {\bar g_{\bar U}}$ on $U \cap \bar U$. \\
2.	${g_{tt}} = {\bar g_{\bar t\bar t}}$, ${g_{tr}} = {\bar g_{\bar t\bar r}}$, ${g_{rr}} = {\bar g_{\bar r\bar r}}$ and $Q = \bar Q$ on $R \times (0,\infty )$. 
\end{lemma}
Lemma \ref{lem14} establishes a one-to-one correspondence between $\operatorname{SO} (3)$-invariant $(0,2)$-tensor fields on $X$ and the quadruples of functions $({g_{tt}},{g_{tr}},P,Q)$, defined on the set $R \times (0,\infty )$. 
As in Theorem \ref{th9}, we have the following observation, in which we consider the unit sphere ${S^2}$ with its canonical metric field. 

\begin{theorem} \label{th15}  
For any $\operatorname{SO} (3)$-invariant metric field $g$ on $X$, there exists a unique metric field ${g_0}$ on $R \times (0,\infty )$, such that $X$ is a warped product of $R \times (0,\infty )$ and ${S^2}$. 
\end{theorem}

\subsection{Isothermal coordinates}
It follows from Lemma \ref{lem14} that the manifold $X = R \times ({R^3}\backslash \left\{ {(0,0,0)} \right\})R \times (0,\infty ) \times {S^2}$, endowed with an $\operatorname{SO} (3)$-invariant metric field $g$, can be considered as the {\it warped product} of two manifolds $R \times (0,\infty )$ (or ${R^2}$)  and ${S^2}$. 
The metric field $g$ expressed by (\ref{eq48}), is induced by the metric fields 
\begin{eqnarray}
&&{g_1} = {g_{tt}}(t,r)dt \otimes dt + {g_{tr}}(t,r)(dt \otimes dr + dr \otimes dt) + {g_{rr}}(t,r)dr \otimes dr, \nonumber \\
&&{g_2} = d\vartheta  \otimes d\vartheta  + {\sin ^2}\vartheta d\varphi  \otimes d\varphi   \label{eq51}
\end{eqnarray}
on $R \times (0,\infty )$ and ${S^2}$ by the function $Q:R \times (0,\infty ) \to R$. 

This property of $\operatorname{SO} (3)$-invariant metric fields allows us to use in our further analysis a classical result on the structure of metric fields $h$ on a 2-dimensional manifold $M$, known as the {\it Korn-Lichtenstein theorem} (Tanaka, Krupka~\cite{ref15}). 
We call a chart $(W,\chi )$, $\chi  = (u,v)$, on $M$ an {\it isothermal chart} for a metric field $h$, if $h$ has an expression 
\begin{eqnarray}
h = f(u,v)(du \otimes du \pm dv \otimes dv).			\label{eq52}
\end{eqnarray}
\begin{lemma} \label{lem16}  
Let $h$ be a metric field on a 2-dimensional manifold $M$. \\
1.If $h$ is a Riemann metric of the H\"older class ${C^{1,0}}$, then each point of $M$ has an isothermal chart. \\
2.	If $h$ is a Lorentz metric of class ${C^1}$, then each point of $M$ has an isothermal chart.
\end{lemma}

\subsection{$\operatorname{SO} (3)$-invariant affine connections}
As in the 3D case, in order to characterize $\operatorname{SO} (3)$-invariant affine connections on $X$, we apply invariance equations (\ref{eq26}). 
We get the following result: 

\begin{lemma} \label{lem17}  
Every $\operatorname{SO} (3)$-invariant connection $\nabla $ on $R \times ({R^3}\backslash \left\{ {(0,0,0)} \right\})$ has the components 
\begin{eqnarray}
&&\nabla _{00}^0 = B_{00}^0(t,r),\quad \nabla _{01}^0 = B_{01}^0(t,r),\quad \nabla _{02}^0 = 0,\quad \nabla _{03}^0 = 0, \nonumber \\
&&\quad \nabla _{11}^0 = B_{11}^0(t,r),\quad \nabla _{12}^0 = 0,\quad \nabla _{13}^0 = 0,\quad \nabla _{22}^0 = {\sin ^2}\vartheta  \cdot B_{22}^0(t,r), \nonumber \\
&&\quad \nabla _{23}^0 = 0,\quad \nabla _{33}^0 = B_{22}^0(t,r), \nonumber \\
&&\nabla _{00}^1 = B_{00}^1(t,r),\quad \nabla _{01}^1 = B_{01}^1(t,r),\quad \nabla _{02}^1 = 0\quad \nabla _{03}^1 = 0,\nonumber  \\
&&\quad \nabla _{11}^1 = B_{11}^1(t,r),\quad \nabla _{12}^1 = 0,\quad \nabla _{13}^1 = 0,\quad \nabla _{22}^1 = B_{22}^1(t,r),\nonumber  \\
&&\quad \nabla _{23}^1 = 0,\quad \nabla _{33}^1 = {\sin ^2}\vartheta  \cdot B_{22}^1(t,r), \nonumber \\
&&\nabla _{00}^2 = 0,\quad \nabla _{01}^2 = 0,\quad \nabla _{02}^2 = B_{02}^2(t,r),\quad \nabla _{03}^2 = \sin \vartheta  \cdot B_{03}^2(t,r), \nonumber \\
&&\quad \nabla _{11}^2 = 0,\quad \nabla _{12}^2 = B_{12}^2(t,r),\quad \nabla _{13}^2 = \sin \vartheta  \cdot B_{13}^2(t,r),\quad \nabla _{22}^2 = 0, \nonumber \\
&&\quad \nabla _{23}^2 = 0,\quad \nabla _{33}^2 =  - \cos \vartheta \sin \vartheta , \nonumber \\
&&\nabla _{00}^3 = 0,\quad \nabla _{01}^3 = 0,\quad \nabla _{02}^3 = \frac{1}{{\sin \vartheta }}B_{03}^2(t,r),\nonumber  \\
&&\quad \nabla _{03}^3 = B_{02}^2(t,r),\quad \nabla _{11}^3 = 0,\quad \nabla _{12}^3 =  - \frac{1}{{\sin \vartheta }}B_{13}^2(t,r),\quad \nabla _{13}^3 = B_{12}^2(t,r), \nonumber \\
&&\quad \nabla _{22}^3 = 0,\quad \nabla _{23}^3 = \cot \vartheta ,\quad \nabla _{33}^3 = 0,  \label{eq53} 
\end{eqnarray}
where $B_{00}^0,B_{01}^0,B_{11}^0,B_{22}^0,B_{00}^1,B_{01}^1B_{11}^1,B_{22}^1,B_{02}^2,B_{03}^2,B_{12}^2,B_{13}^2$ are arbitrary functions of the variables $t$ and $r$. 
\end{lemma}

\subsection{$\operatorname{SO} (3)$-metrizability}
The following two theorems give the solution to the $\operatorname{SO} (3)$-metrizability problem for the manifold $X = R \times ({R^3}\backslash \left\{ {(0,0,0)} \right\})$. 

\begin{theorem} \label{th18}  
Let $\nabla $ be an affine connection on $X$. The following conditions are equivalent: \\
1.	$\nabla $ is $\operatorname{SO} (3)$-invariant and $\operatorname{SO} (3)$-metrizable. \\
2.	The non-zero components of $\nabla $ satisfy 
\begin{eqnarray}
&&\mp {\nabla ^0}_{11} = {\nabla ^1}_{01} = {\nabla ^0}_{00} = {A^0}_{00}(u),\quad {\nabla ^0}_{01} =  \mp {\nabla ^1}_{00} = {\nabla ^1}_{11} = {A^1}_{11}(v), \nonumber \\
&&\nabla _{22}^1 =  \mp \frac{{{C_2}}}{{{C_1}}}{A^2}_{12}(v){\text{exp}}\left( {2\int_1^v {A_{12}^2(V) - A_{11}^1(V)dV} } \right){\text{exp}}\left( { - 2\int_1^u {A_{00}^0(U)dU} } \right),\nonumber  \\
&&\nabla _{33}^1 = \nabla _{22}^1{\sin ^2}\vartheta ,\quad \nabla _{12}^2 = A_{12}^2(v),\quad \nabla _{33}^2 =  - \sin \vartheta \cos \vartheta ,\quad \nabla _{23}^3 = \cot \vartheta ,  \label{eq54} 
\end{eqnarray}
where the sign $ \pm $ stands for upper case as Riemann, and bottom as Lorentz, 
${C_1},{C_2} \in R$ are nonzero constants, and $A_{00}^0 = {A^0}_{00}(u)$, $A_{11}^1 = {A^1}_{11}(v)$,$A_{12}^2 = A_{12}^2(v)$ are arbitrary functions. 
\end{theorem}

We can now formulate our main result in this section, namely an explicit description of solutions of the $\operatorname{SO} (3)$-metrizability problem on the manifold $X = R \times {R^3}\backslash \left\{ {(0,0,0)} \right\}$. 

\begin{theorem} \label{th19}  
If an $\operatorname{SO} (3)$-invariant affine connection $\nabla $ satisfies the local metrizability conditions (\ref{eq54}), then the $\operatorname{SO} (3)$-metrizability problem has a solution 
\begin{eqnarray}
g = P(u,v)(du \otimes du \pm dv \otimes dv) + Q(v)(d\vartheta  \otimes d\vartheta  + {\sin ^2}\vartheta d\varphi  \otimes d\varphi ), \label{eq55}
\end{eqnarray}
where 
\begin{eqnarray}
  P(u,v) = {C_1}{\text{exp}}\left( {2\int_1^u {A_{00}^0(U)dU} } \right){\text{exp}}\left( {2\int_1^v {A_{11}^1(V)dV} } \right), \nonumber \\
  Q(v) = {C_2}{\text{exp}}\left( {2\int_1^v {A_{12}^2(V)dV} } \right),\quad {C_1},{C_2} \in R,  \label{eq56}
\end{eqnarray}
and the sign $ \pm $ stands for upper case as Riemann, and bottom as Lorentz. 
\end{theorem}

We can apply to this result Lemma \ref{lem14}; we conclude that formula (\ref{eq55}) defines a global metric field on the manifold $X$.

\section*{Acknowledgments}  
The authors acknowledge the support of the Czech Science Foundation (grant 201/09/0981), the Czech Ministry of Education, Youth, and Sports (grant MEB 041005), National Science Foundation of China (grant No. 109320020) and Palacky University (grant No. PrF-2010-008, PrF-2011-022). The first author thanks YITP of Kyoto University. The second author also appreciates the support of the Department of Theoretical Physics and Astrophysics, Masaryk University in Brno, and fruitful discussions with the seminar members.


\section*{References}


\begin{thebibliography}{0}
\bibitem{ref1} M. Anastasiei, Metrizable linear connections in vector bundles, Pub. Math. Debrecen {\bf 63} (2003) 277-288.

\bibitem{ref2} M. Crampin, On the differential geometry of the Euler-Lagrange equations, and the inverse problem of Lagrangian dynamics, J. Phys.A, Math. Gen. {\bf 14} (1981) 2567-2575

\bibitem{ref3} M. Crampin, G.E. Prince, G. Thompson, A geometrical version of the Helmholtz conditions in time-dependent Lagrangian dynamics, J. Phys.A, Math. Gen. {\bf 17} (1984) 1437-1447

\bibitem{ref4} J. Dieudonne, {\it Foundations of Modern Analysis}, Academic Press, New York and London, 1969

\bibitem{ref5} L.P. Eisenhart, O. Veblen, The Riemannian geometry and its generalization, Proc. Nat. Acad. Sci. {\bf 8} (1922) 19-23 

\bibitem{ref6} S. Helgason, {\it Differential Geometry and Symmetric Spaces}, Academic Press, New York, 1978 

\bibitem{ref7} O. Kowalski, Metrizability of affine connections on analytic manifolds, Note di Matematica {\bf 8} (1998) 1-11

\bibitem{ref8} O. Kowalski, On regular curvature structures, Math. Z. {\bf 125} (1972) 129-138

\bibitem{ref9} D. Krupka, A. Sattarov, The inverse problem of the calculus of variations for Finsler structures, Math. Slovaca {\bf 35} (1985) 217-222

\bibitem{ref10} D. Krupka, O. Stepankova, On the Hamilton form in second order calculus of variations, in: Proc. Internat. Meeting "Geometry and Physics", Florence, October 1982, Pitagora, Bologna, 1983, 85-101 

\bibitem{ref11} G.E. Prince, M. Crampin, Projective differential geometry and geodesic conservation laws in general relativity. I: Projective actions, General Relativity and Gravitation {\bf 16} (1984) 921-942

\bibitem{ref12} W. Sarlet, The Helmholtz conditions revisited. A new approach to the inverse problem of Lagrangian dynamics, J. Phys. A, Math. Gen. 15 (1982) 1437-1447 

\bibitem{ref13} B.G. Schmidt, Conditions on a connection to be a metric connection, Commun. Math. Phys. {\bf 29} (1973) 55-59 

\bibitem{ref14} L. Tamassy, Metrizability of affine connections, Balcan J. Geom. Appl. {\bf 2} (1997) 131-138

\bibitem{ref15} E. Tanaka, D. Krupka, Isothermal coordinates: Elementary existence proof and applications, Seminar Global Analysis and Applications, Masaryk University in Brno, 2011; to appear

\bibitem{ref16} Z. Vilimova, Metrizability problem of a linear connection (in Czech), Thesis, Silesian University Opava, 2003

\end{thebibliography}
\end{document}